# Optimization of a three-phase Induction Motor for Electric Vehicles Based on Hook-Jews Optimization Method

1st A. Mousaei
Dept. Electrical and Computer Engineering
University of Tabriz
Tabriz, Iran
a.mousaei@tabrizu.ac.ir

2nd S. A. Mohammadabadi
Dept. Electrical and Computer Engineering
University of Pisa
Pisa, Italy
Azimisah@gmail.com

*Abstract*— In this paper, the Hook-Jews (HJ) optimization method is used to optimize a 3-phase Squirrel-Cage Induction Motor (SCIM) as an Electric Vehicle's (EV) motor. Optimal designs with different numbers of poles, different nominal and maximum speeds, and different numbers of grooves are compared and the best one is selected. The optimization method used has advantages such as simple programming, omission of gradients, short convergence times, and the possibility of changing individual parameters. Design parameter variations for optimal designs with rated speeds for 2-pole and 4-pole motors are shown and explained. The results show that his 2-pole motor with the rectangular stator and rotor slots and a rated speed of 1800 rpm has the highest efficiency.

## I. INTRODUCTION

Car traffic is the main cause of respiratory and noise pollution in big cities. Using Electric Vehicles (EVs) can be the solution to these two problems. Among the advantages of EVs is the possibility of accessing electric energy through the electric energy distribution system. The big disadvantage of these cars is the low power density and long battery charging time. Therefore, in addition to proper vehicle control and proper energy management, optimal design of its various parts, in particular the motor, is also necessary.

The topic of EVs goes back to the beginning of the 20th century (around 1916). But due to the abundance of fossil energy, the lack of awareness of the problem of air pollution, and the limitation of electrical energy resources, it was forgotten for a long time. Around the year 1980, the crisis of fossil energies due to their reduction and the sharp increase in the price of fuel, and noise and respiratory air pollution caused attention to EVs again. In this way, different EVs with different Electric Motors (EMs) were designed and built.

Among EMs, DC motors are considered for their simple control system, and Switch Reluctance Motors (SRMs) for their simple construction and easy control. The problems caused by the commutator and brush system of DC motors and the high noise of SR motors caused more attention to AC motors. In recent years, Induction Motors (IMs) are very popular due to their simple construction, low price, and high energy density.

The main requirements of EV's motors are: [2] torque to-inertia ratio and high power-to-weight ratio, high breaking torque (up to 400 percent), high speed, low noise level, no need for maintenance, small size and lightweight, simple control Reasonable price, high efficiency with different speeds (low and high) and the ability to return energy when braking or moving downhill. It should be mentioned that the squirrel cage induction motor has more of these requirements.

Different optimization methods have been used for the optimal design of induction motors [3] to [16]. These methods are Random Search (RS), Direct Search (DS) or Hook-Jewes (HJ), Simple (S), Pavel (P), Daviden-Fletcher-Pavel (DFP) method, Deepest Slope (DS), direct search based on the Rosenberg method, First-Order Gradient (FOG), Monica (M), Niching Genetic Algorithm and optimization based on Neural Network (NN).

TABLE I. COMPARISON OF OPTIMAL DESIGN RESULTS OF FIVE DIFFERENT METHODS

| Method / Parameter | DS | DFP | P | HJ | RS |
|---|---|---|---|---|---|
| *Braking torque (pu)* | 3.01 | 2.7 | 3.05 | 3.12 | 3.32 |
| *Starting torque (pu)* | 1.85 | 1.51 | 1.91 | 1.95 | 2.23 |
| *Starting current (A)* | 51.7 | 48.5 | 50.9 | 52 | 53.2 |
| *Power factor* | 0.85 | 0.88 | 0.85 | 0.85 | 0.855 |
| *Slip (%)* | 3.45 | 3.22 | 3.61 | 3.5 | 3.83 |
| *Increase in temperature (°C)* | 64.3 | 72.5 | 74.96 | 74.05 | 63.3 |
| *Price* | 331.05 | 292.53 | 265.58 | 263.8 | 325.97 |
| *Convergence time (%)* | 100 | 50 | 50 | 15 | 35 |
| *Number of repetitions* | 11 | 30 | 4 | 3 | 9 |

Table I compares the results obtained from five different optimization methods with the objective function of the price of consumables, in the case of a 7.5 kW, IM with four poles [6] and [7].

The comparison of the results in Table I shows that the HJ method has a better performance. Based on this, and also the recommendations of various authors [3] to [5], and [10], the HJ method has been chosen for the optimal design of the EV's IM.

In addition to the dimensional parameters, the efficiency of an EV depends on its nominal values and fixed parameters. These items include power, voltage, nominal speed and torque, maximum speed and maximum torque, number of poles, and shape, and number of grooves. Nominal power, maximum speed, nominal torque, and maximum torque are determined based on the intended efficiency of the EV. The system's voltage is selected based on the available possibilities. Choosing a higher voltage provides the possibility of better performance, however, some problems such as the limitation of the voltage of the batteries limit this possibility. The number of grooves can be calculated based on design standards. The nominal speed, the number of poles, and the shape of the grooves should be examined more precisely. Based on this, it is the best design that can be obtained.

In this article, the effect of the number of poles, nominal speed, and the shape of the grooves on the efficiency of the three-phase EV's IM is studied and evaluated. The 15-hp three-phase IM for EV is optimally designed with different values of these parameters and finally, the final optimal design is introduced. The article consists of six sections. In the second part, the effect of the number of poles, nominal speed, and the shape of the grooves on the efficiency of the motor is studied. In the third section, the method of calculating engine efficiency is presented. In the fourth section, the optimal design is studied. In the fifth section, a 15 hp sample engine is designed and the effects of the abovementioned parameters are discussed and investigated. Finally, the conclusion is made in the last section.

## II. THE EFFECT OF THE NUMBER OF POLES, GROOVES TYPE, AND NOMINAL SPEED ON THE EFFICIENCY OF THE MOTOR

The EV's IM must work in a wide range of speeds (zero to several thousand rotations per minute). In addition, other requirements of EVs such as sufficient torque in the acceleration mode or working at a constant speed, small mechanical time constant (dynamic and fast), and so on, should be provided by IM. Due to the limitation of battery energy, the available energy should be used in the best way, so it should have a high efficiency and its design should be such that the torque fluctuations (caused by the harmonics of the power source) are reduced within this range. and have little noise. In order to have fast dynamics, the system should be as light as possible. Of course, you can consider the goal of optimal volume to occupy the least space. Therefore, its design is different from the design of standard IMs to limit the industry. The current in the Voltage Source Inverter (VSI) supply should be high leakage inductance, but not so much that it enters the unstable operating range at different motor speeds. To reduce the keying frequency and ultimately cost less, and to reduce keying losses, consider reducing the effect on the skin and related losses, deep grooves should not be used. Because the motor with an inverter power supply will not have the problem of starting torque, there is no need to use the deep grooves of the rotor [13], [17] to [20] in order to increase the acceleration during acceleration and also the acceleration during positive acceleration. The nominal values of the elements of the power converter (inverter), and the radius of the rotor should be as small as possible, and instead of its length, it should be selected as large. Because the power supply voltage range is limited due to the use of a battery, therefore, for the specific power of the current and the thickness of the conductors, a large groove space is necessary. Considering the high speed of the motor, the diameter of the rotor must be small, so the number of grooves on the pole is small. In this order, the number of motor poles is chosen between 2 and 4 [13] and [20].

Considering the high frequency in the motor, it is possible to reduce the skin effect and the losses caused by it by choosing the appropriate shape of the rotor groove. For this reason, usually in high-speed IMs (such as EV) the shape of the rotor groove is rectangular its length and width are very close to each other. In addition, the voltage can be increased to better distribute the flux in different parts of the groove and reduce the harmonic losses.

The maximum speed of the motor depends on the selected power transmission system and the maximum speed of the vehicle, but the nominal and base speed (the base speed is selected equal to the nominal speed), the pole, the type of groove, the losses and other conditions of the motor's performance and the operating conditions are also used in the selection process. Choosing the nominal speed with regard to the working frequency of the machine at this speed has a direct effect on the keying losses and harmonics. On the other hand, it will also be effective on the moment of inertia of the motor, so choosing the optimal nominal speed can be done at the same time. It provided transients for the motor and ultimately for the EV.

## III. CALCULATION OF EFFICIENCY CHARACTERISTICS

In the inverter feeding of the IM, in addition to the first (main) harmonic, individual harmonics also appear, and hence the efficiency characteristics of this motor are different from those with sinusoidal feeding. The most important losses in the IM with inverter feeding can be classified and calculated as follows (of course, there are other losses that are considered in the design software, but they are mentioned here for the sake of simplicity).

### A. Core losses

The total core losses in an IM's inverter can be calculated using equation 1:

$$P_c = \sum_m P_{cm} = \sum_m (P_{hm} + P_{em}) \tag{1}$$

Each of Foucault's Current and hysteresis of harmonics is firstly calculated in the unit of weight of different parts of the magnetic circuit of the motor (core) according to the following relations and then by multiplying the weight of the relevant part, the harmonic losses of each of these parts are calculated separately for each of the harmonics:

$$P_{hm} = \sum_i P_{hmi} = \sum_i G_i P_{hmi} \tag{2}$$

$$= \sum_i G_i K_h \sigma_h f_m B_{mmi}^k$$

$$P_{em} = \sum_i P_{emi} = \sum_i G_i P_{emi} \tag{3}$$

$$= \sum_i G_i K_e t^2 f_m^2 B_{mmi}^2 K_{EM/pi}$$

### B. Resistor losses of stator and rotor

The total losses of the stator and rotor can be calculated from the sum of the losses related to the different current harmonics in the form of the following relationship:

$$P_\Omega = \sum_m 3(R_{sm} I_{sm}^2 + R_{rm} I_{rm}^2) \tag{4}$$

## C. Mechanical losses (friction, ventilation and air resistance)

With the help of the following relationship, these losses can be calculated at different speeds:

$$P_{fw} = 8D_r(L + 0.15)v_a^2 \tag{5}$$

Finally, all the losses and efficiency are:

$$P_{loss} = P_\Omega + P_c + P_{fw} + (P_p + P_K + P_z + P_{bll}) \tag{6}$$

$$P_{in} = \sum_m 3V_{sm}I_{sm}\cos\varphi_m \tag{7}$$

$$\eta = (P_{in} - P_{loss})/P_{in} \tag{8}$$

## D. Nominal torque and failure torques at nominal and maximum speeds

The following relations obtain the values of these torques:

$$T_n = \sum_m 1.5PR_{rm}I_{rm}^2/(mf_m s_m) \tag{9}$$

$$T_{pb} = 1.5E_{s1}/(X_{r1}\omega_s) = T_n R_{r1}/X_{r1} \tag{10}$$

$$T_{pm} = (f_b/f_{max})^2 T_{pb} \tag{11}$$

## E. Inertia coefficient

This coefficient, which determines the positive and negative accelerations of the motor, is calculated as follows.

$$H = 0.5J\omega_r^2/Q \tag{12}$$

### IV. OPTIMUM DESIGN OF MOTOR

Various methods have been presented for the optimal design of IM [3] to [16], which are used for the optimal design of EV's IM. In this article, the HJ method has been used for this purpose [3] to [5] and [10] to [12]. In this section, only the objective functions, limitations, and optimization variables of EV's IM are mentioned.

There are different points of view on the optimal design of a motor, which are: Minimum cost, maximum efficiency, minimum volume or weight, good performance (such as low slippage, high power factor, etc.), and a multifaceted view (combination of different views). In an EV, considering the limitation of energy and battery power should be used of them, so, efficiency is definitely important. In addition, in order to reduce the weight of the whole car and in other words, to reduce the energy consumption, the motor should be lightweight, so it seems that in the case of the optimal design of the EV, the multi-faceted point of view is weight and efficiency. The objective function can be chosen. Of course, it is viable to make a design with different combinations of these points of view and choose the best one that leads to the most suitable expected result as the optimal design. In this article, considering that the main goal is to compare the effect of choosing the number of poles, the type of groove, and the nominal speed on the stable and transient performance of the motor, only the efficiency function has been discussed.

In the design optimization, other goals are also pursued, which as improving or at least maintaining the desired performance for the motor, which can be mentioned as secondary goals. Since enlarging the objective function (a multidimensional function with various and many components) causes the optimization process to be slow and the scope of the search for the optimized software is limited, and probably not very favorable results will be obtained. Therefore, these secondary objectives are not directly included in the objective function but are applied as constraints to the optimization process.

The most important restrictions considered in this article: minimum power factor (0.85), the maximum increase in temperature (75 °C), minimum production torque, the minimum ratio of failure torque to nominal torque at nominal speed (1.5), minimum breakdown torque at maximum speed (3.5 Nm), maximum rotor speed (120 m/s), maximum rotor time constant (4 s), the maximum flux density of stator's tooth (1.2 T), maximum total cost (if necessary), and maximum weight or volume (if it is not part of the target function).

There are many design variables of IM. Often, the more the number of optimization variables cause the better optimization result but the speed of convergence is greatly reduced and it will be difficult to control the optimization process. Therefore, we try as much as possible to prevent the increase of the number of variables and also to use variables that have a greater effect on the optimal design as the main optimization variables. Based on this, in this article, these variables are: the inner diameter of the stator, length of the core, width and depth of the stator groove, width and depth of the rotor groove, the depth of the stator and rotor rings, the length of the air gap, the cross-sectional surface of the rings on the outer part of the rotor, and the airgap's flux density.

In addition to these variables, other variables including working voltage, nominal and maximum speeds, the number of stator and rotor grooves, the type of stator and rotor grooves, and the number of poles can also be considered. In this article, the working voltage of the motor has been chosen to be equal to 96 V due to the limitation of the battery voltage and other restrictions. The maximum speed of the motor has been selected according to the maximum speed requested by the 900 rpm. The number of slots of the stator and rotor with the change and repetition of optimization are 18 and 13 for the 2-pole motor, and 24 and 18 for the 4-pole motor, respectively. The nominal speed, and the type of grooves and the number of poles are the things that are discussed more in the article, and different optimal designs with different values of the nominal speed, the type of groove and the number of poles are examined and their efficiency is compared. and finally, based on the comparisons, the best design is presented.

### V. OPTIMAL DESIGN OF THE SAMPLE MOTOR

In this part, a squirrel cage IM for EV is designed and the effect of choosing the number of poles, the type of stator and rotor slots, and the nominal speed on the performance characteristics of the motor is examined and analyzed. In the design, the first five harmonics of the supply voltage (5th, 7th, 11th, 13th, and 17th harmonics) are taken into account with values of 0.972, 0.088, 0.019, 0.015, and 0.050, respectively. The nominal motor's power is 15 hp and the maximum speed of the motor is selected at 900 rpm. According to the comparisons in section II, the choice of the number of poles will be limited to 2 and 4. To compare the effect of the grooves, two types of rectangular and round grooves are studied in the limit state. Considering that increasing the nominal speed leads to a decrease in the efficiency of the system, the nominal speed range is chosen from 1600 rpm to 2000 rpm. The number of slots of the stator and rotor in the case of the 2-pole design is 18 and 13 respectively, and in the case of the 4-pole design, choose 18 and 4 more. Tables 2 and 3 show the results of different optimal designs for stator and rectangular rotor grooves, rectangular stator grooves, and round rotor, respectively. It is observed that for rectangular stator and rotor grooves:

- The volume of the 2-pole motor for three nominal speeds is significantly less than that of the 4-pole motor (43 % on average).
- The inner diameter of the stator of the 2-pole motor is smaller, and on the contrary, the outer diameter of the stator is larger than that of the 4-pole motor.
- Contrary to the larger volume of the four-pole motor, its weight is less (rectangular and the groove of the rotor is round).
- The moment of inertia of the rotor of the 2-pole motor is significantly less than that of the 4-pole motor (49 % on average).
- The efficiency of the two-pole motor is 0.96 % higher than the four-pole motor. The average power factor of the 2-pole motor is 17 % higher than that of the 4-pole motor, and the power factor is not met in the 4-pole motor, but it is met in the 2-pole motor.
- The braking torque at the maximum speed of the 2-pole motor is higher (2.85 % on average), so the 2-pole motor can have more overload at the maximum speed.
- The braking torque at the nominal speed of the 2-pole motor is on average 3.04 % higher, so the 2-pole motor can have a more additional load at the nominal speed and also more dynamic acceleration.
- The increase in temperature has been calculated in all 2-pole designs, but in the case of 4-pole designs, it has only been calculated for the nominal speed of 1600 rpm.
- The price of 2-pole designs is higher (on average 14.35 %).

Almost similar results can be obtained in the case of designs with rectangular grooves for the stator and round for the rotor (Table III). In general, it can be concluded that the designs with rectangular stator and rotor grooves are better in terms of performance compared to rectangular stator grooves and round rotor grooves.

TABLE II.  OPTIMAL DESIGN RESULTS FOR RECTANGULAR STATOR AND ROTOR GROOVES

| P | 2 | | | 4 | | |
|---|---|---|---|---|---|---|
| Nominal Speed (rpm) | 1600 | 1800 | 2000 | 1600 | 1800 | 2000 |
| L (m) | 0.0849 | 0.0854 | 0.0781 | 0.1398 | 0.1326 | 0.1301 |
| $D_0$ (m) | 0.2234 | 0.2203 | 0.2241 | 0.2159 | 0.2076 | 0.2050 |
| D (m) | 0.1288 | 0.1264 | 0.1293 | 0.1398 | 0.1326 | 0.1303 |
| W (kg) | 40.68 | 39.69 | 38.67 | 37.78 | 33.75 | 32.45 |
| V ($m^3$) | 0.0033 | 0.0033 | 0.0031 | 0.0051 | 0.0045 | 0.0043 |
| C ($) | 125.0 | 122.2 | 119.3 | 112.9 | 102.3 | 98.7 |
| J ($kgm^2$) | 0.0287 | 0.0265 | 0.0272 | 0.0477 | 0.0369 | 0.0338 |
| η (%) | 86.00 | 85.85 | 85.50 | 85.47 | 84.99 | 84.43 |
| pf | 0.865 | 0.860 | 0.857 | 0.741 | 0.735 | 0.729 |
| T (°C) | 73.97 | 71.83 | 67.22 | 71.83 | 77.08 | 77.86 |
| $T_{pm}$ (Nm) | 3.495 | 3.759 | 4.042 | 3.335 | 3.677 | 3.962 |
| $T_{pb}$ (Nm) | 110.57 | 93.97 | 81.85 | 105.52 | 91.93 | 80.24 |
| $W_s$ (m) | 0.0111 | 0.0111 | 0.0111 | 0.0111 | 0.0111 | 0.0111 |
| $d_s$ (m) | 0.0267 | 0.0267 | 0.0267 | 0.0261 | 0.0261 | 0.0261 |
| $W_r$ (m) | 0.0149 | 0.0146 | 0.0144 | 0.0127 | 0.0124 | 0.0122 |
| $d_r$ (m) | 0.0238 | 0.0234 | 0.0246 | 0.0181 | 0.0153 | 0.0149 |

TABLE III.  OPTIMUM DESIGN RESULTS FOR RECTANGULAR STATOR GROOVES AND ROUND ROTOR GROOVES WITH A NOMINAL SPEED OF 1800 RPM

| P | 2 | 4 | P | 2 | 4 | P | 2 | 4 |
|---|---|---|---|---|---|---|---|---|
| L (m) | 0.1000 | 0.0844 | $D_0$ (m) | 0.3621 | 0.2283 | D (m) | 0.1264 | 0.1398 |
| W (kg) | 43.06 | 29.49 | V ($m^3$) | 0.0038 | 0.0035 | $C_t$ ($) | 131.5 | 89.2 |
| J ($kgm^2$) | 0.0294 | 0.0308 | η (%) | 85.07 | 85.24 | pf | 0.853 | 0.741 |
| T (°C) | 73.81 | 79.53 | $T_{pm}$ (Nm) | 4.149 | 3.612 | $T_{pb}$ (Nm) | 103.72 | 90.30 |
| $W_s$ (m) | 0.0111 | 0.0090 | $d_s$ (m) | 0.0267 | 0.0331 | $W_r=d_r$ (m) | 0/0175 | 0.0149 |

In order to compare more and better the effect of the nominal speed on the performance of the motors, a number of parameters in Table II have been drawn as a curve according to the nominal speed. Figures 1A to 1E, in the order of the curve of changes in stator inner diameter, the core length, core volume, core weight, and windings, moment of inertia, value, breaking torque at maximum and nominal speeds, efficiency and power factor with the change of speed from 1400 rpm to 2800 rpm for 2-pole designs, they show. Figures 2A to 2E show the same curves for 4-pole designs in the nominal speed range of 1400 rpm to 2100 rpm. In the case of both 2-pole and 4-pole designs, the increase in the nominal speed on average causes a decrease in the length of the core, inner diameter of the stator, volume and weight, moment of inertia, cost, breaking torque at the nominal speed, efficiency and power factor, and vice versa for the breaking torque in the braking torque. It increases by approximately. Considering that the objective function is efficiency optimization, so it can be said that increasing the nominal speed causes the design quality to decrease, or in other words, reducing the nominal speed causes the design to improve. From the point of view of increasing the nominal speed from the point of view of volume, weight, moment of inertia and the desired price, these things are very important in the selection of motors for electric vehicles. So, an option should be selected according to the above requirements and the intended prioritization in the case of the engine, so that all the main goals and secondary goals are met as much as possible.

In the optimal design of the motor, only the losses of the motor itself are taken into account and the switching losses are not taken into account. Whenever the switching losses are also taken into account, due to the fact that increasing the nominal speed will increase the switching losses and increase the volume and nominal values of the power elements, this increase in the nominal speed is more than the nominal speed.

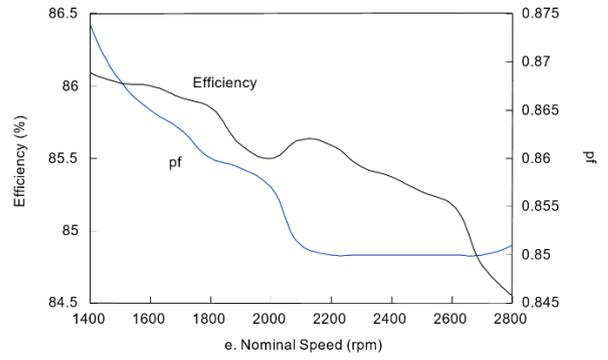

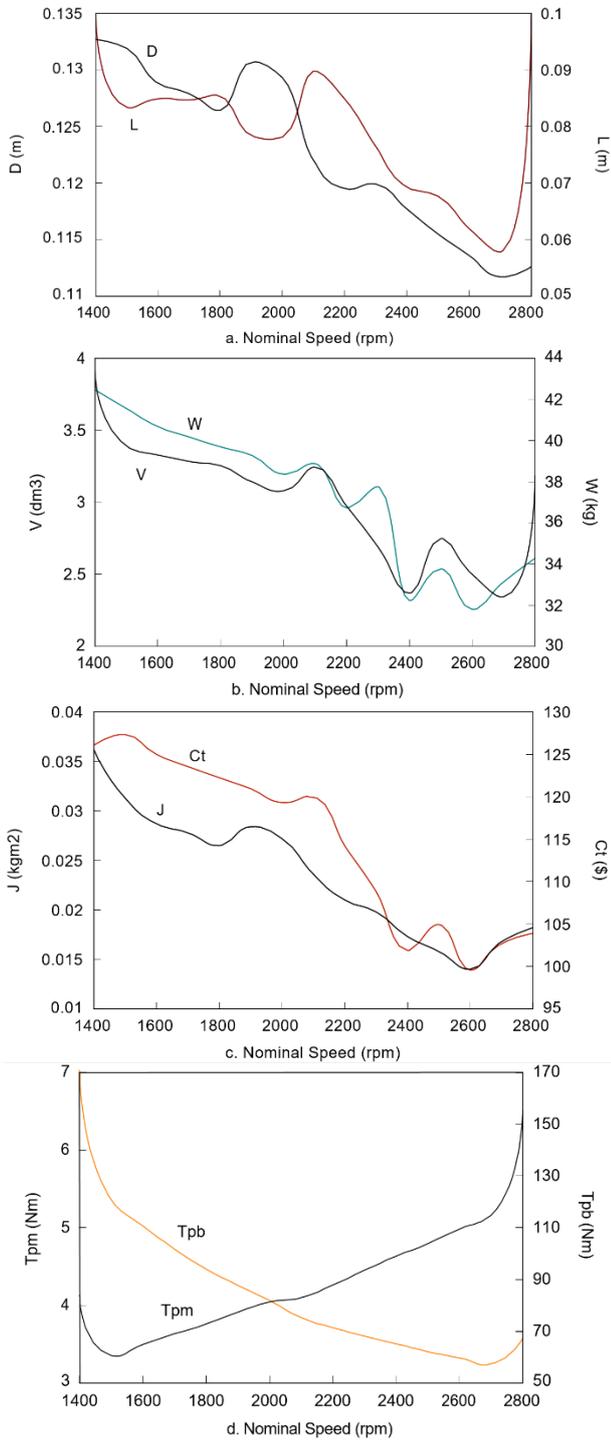

Fig. 1. Changes in the characteristics and main parameters of the motor for 2-pole designs with an increase in the nominal speed

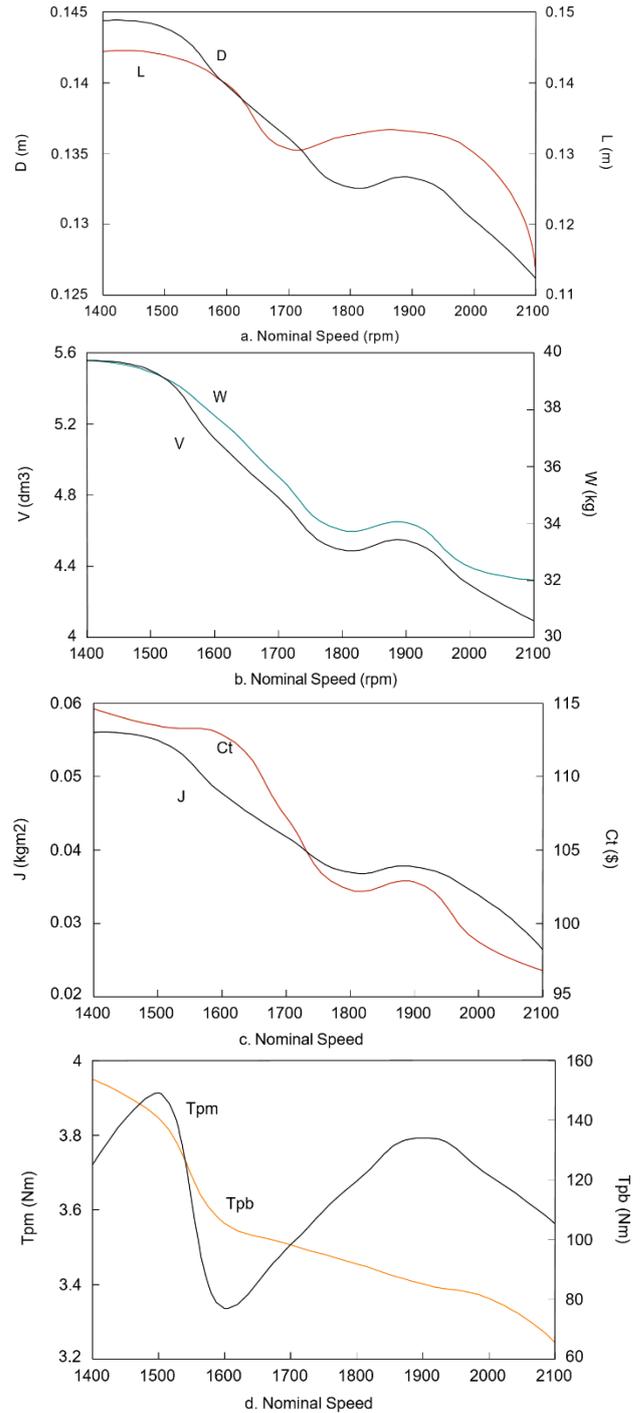

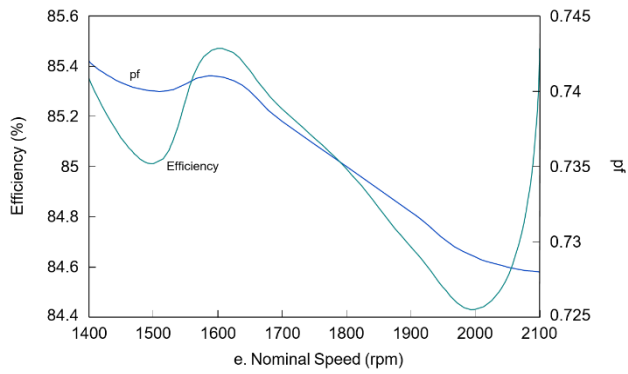

Fig. 2. Changes in the characteristics and main parameters of the motor for 4-pole designs with an increase in the nominal speed

## VI. CONCLOTION

Examining and analyzing different plans with different combinations gives the following results:

i. Despite choosing a low supply voltage (in order to limit the voltage of the batteries) and considering the five harmonics in the voltage waveform, optimal design results are desirable.

ii. The advantages of the 2-pole motor compared to the 4-pole motor are: small size and moment of inertia, high efficiency, better power factor, and high breaking torque at maximum and nominal speeds.

iii. The disadvantages of the 2-pole motor compared to the 4-pole motor are: weight and high price.

iv. Rectangular grooves for the stator and rotor in 2-pole designs lead to better motor efficiency compared to rectangular grooves for the stator and round grooves for the rotor. For 4-pole designs, the opposite of the above is true.

v. For 2- and 4-pole designs, increasing the nominal speed in the design phase leads to volume, weight, the moment of inertia, price, breakdown torque at nominal speed, efficiency, and power factor, and vice versa, the breakdown torque increases at maximum speed. Considering the objective function (yield), these results show that increasing the nominal speed leads to an unfavorable plan, while the secondary objectives (volume, weight, and moment of inertia) require that the nominal speed be increased. Another factor to limit the nominal speed.

vi. Summarizing the results of the constant work mode, the 2-pole motor with a nominal speed of 1800 rpm shows the most suitable design for an electric car.

### *List of symptoms*

| Symbol | Description |
|---|---|
| $P_c$, $P_{cm}$ $P_{hm}$, $P_{em}$ | Foucault, hysteresis and core losses to m-th harmonic, and total core losses |
| $P_{hmi}$, $P_{emi}$ | Total Foucault and hysteresis losses and unit weight of i-th part and m-th harmonic |
| $K_h$, $K_e$ | Coefficients of Foucault and hysteresis |
| $\sigma_h$ | For steel sheets with a thickness of 0.35 mm, it is equal to 3. |
| $S_m$, $f_m$ | Its frequency and shift are proportional to the m-th harmonic |
| $B_{mmi}$ | The maximum magnetic flux density corresponding to the m-th harmonic in the i-th part according to Tesla |
| $K$ | For today's magnetic materials, it is about 2 |
| $t$, $\rho_I$ | Specific resistance in ohms per centimeter and thickness of steel sheets in millimeters |
| $G_i$ | The weight of the i-th part is in kilograms |
| $K_{Em}$ | The permeability coefficient for m-th harmonic |
| $P_{fw}$, $P_\Omega$ | Ohmic losses and mechanical losses according to watts |
| $X_{r1}$, $R_{sm}$, $R_{rm}$ | The rotor and stator resistances corresponding to the third harmonic and the leakage reactance of the rotor corresponding to the first harmonic |
| $\cos \varphi_m$ $V_{sm}$, $I_{sm}$ $I_{rm}$ | Rotor currents, stator currents, voltage and power factor of the input motor corresponding to harmonic mam according to ampere and volt |
| $D$, $D_0$ $L$ | The diameter and length of the rotor core, the outer and inner diameters of the stator |
| $D_s$, $d_r$ $W_s$ | Width and depth of rotor and stator grooves |
| $V_a$ | The linear speed of the rotor in meters per second |
| $P_h$ | Losses caused by the leakage flow of teeth according to Watt |
| $P_K$ | Losses caused by the flow of leakage due to the grooves being diagonal |
| $P_Z$ | Losses caused by zigzag leakage flow according to watts |
| $P_{b11}$ | Losses caused by leakage flow of rotor bars according to watts |
| $T_{pb}$, $T_{pm}$ $T_n$ | Nominal torque, failure torques at maximum and nominal speeds |
| $Z_{s1}$, $Z_{t1}$ | Total impedance and stator impedance corresponding to the first harmonic according to Ohm |
| $f_{max}$, $f_b$ | Nominal and maximum frequencies |
| $Pf$, $\eta$ | Efficiency and power factor |
| $C_t$, $V$, $W$ | Weight, volume and price of the core and windings |
| $H$, $J$, $T$ | Increasing the temperature, moment of inertia and inertia constant of the motor |